# An all-optical spatial light modulator for field-programmable silicon photonic circuits

Roman Bruck,[1] Kevin Vynck,[2] Philippe Lalanne,[2] Ben Mills,[3] David J. Thomson,[3] Goran Z. Mashanovich,[3] Graham T. Reed,[3] and Otto L. Muskens[1,*]

[1] *Physics and Astronomy, Faculty of Physical Sciences and Engineering, University of Southampton, Southampton SO17 1BJ, UK*

[2] *LP2N, UMR 5298, CNRS – IOGS – Univ. Bordeaux, 33400 Talence, France*

[3] *Optoelectronics Research Centre, University of Southampton, Southampton SO17 1BJ, UK*

*Corresponding author: o.muskens@soton.ac.uk

**Reconfigurable photonic devices capable of routing the flow of light enable flexible integrated-optic circuits that are not hard-wired but can be externally controlled. Analogous to free-space spatial light modulators, we demonstrate all-optical wavefront shaping in integrated silicon-on-insulator photonic devices by modifying the spatial refractive index profile of the device employing ultraviolet pulsed laser excitation. Applying appropriate excitation patterns grants us full control over the optical transfer function of telecommunication-wavelength light travelling through the device, thus allowing us to redefine its functionalities. As a proof-of-concept, we experimentally demonstrate routing of light between the ports of a multimode interference power splitter with more than 97% total efficiency and negligible losses. Wavefront shaping in integrated photonic circuits provides a conceptually new approach toward achieving highly adaptable and field-programmable photonic circuits with applications in optical testing and data communication.**

## 1. Introduction

With optical links replacing electrical connections on ever shorter length scales, all-optical control of optical signals becomes ever more desirable to circumvent the electronic speed limit and avoid transformation losses between optical and electrical domains. All-optical switches and modulators have been demonstrated based on different integrated optical designs [1-9]. Most of the conventional tuneable photonic devices use one specific parameter to establish precise control over the output(s), but lack the freedom to completely govern the transmission function. For static devices, many new degrees of freedom can be accessed by exploiting the vast design space offered by complex device geometries [10-14]. In fact, defining the spatial refractive index profile in a small planar area is sufficient to create compact devices with arbitrary characteristics, as was recently demonstrated experimentally by etching an optimized pattern into a silicon-on-insulator (SOI) slab[15].

Here we show that spatial refractive index profiles can be controlled all-optically, allowing us to dynamically route light between outputs of a SOI multimode interference (MMI) device. Our approach uses the vast capabilities offered by digital micro-mirror device (DMD) technology in spatial light modulation in conjunction with an ultrafast photomodulation setup [16]. As sketched in Fig. 1(a), a DMD is used to project a pattern of femtosecond laser pulses onto the MMI-device surface, allowing simultaneous modulation of the refractive index in a large number (around 500) of positions. For each illuminated position, plasma dispersion locally decreases the refractive index of silicon by approximately 0.25 refractive index units (see methods), thus achieving a significant perturbation of the light flow in the device.

Wavefront shaping has been demonstrated as a powerful tool in both characterizing and controlling the optical modes in three-dimensional complex media [17-19]. Translating these capabilities to an integrated platform opens up significant opportunities for freeform shaping of optical signals in integrated photonic circuits. We specifically demonstrate this general technique on the example of shaping wavefronts in SOI MMI regions effectively and with negligible loss. The transfer function of MMI-devices is defined by the interference of a large number of transverse modes. Any localized perturbation in the refractive index profile introduces coupling between the MMI modes (see Fig. 1(b)) and allows the redistribution of power between modes. Considering a single perturbation, modelled by a refractive index change $\Delta n = -0.25$ over a square 700 nm x 700 nm pixel, we find with rigorous coupled-mode simulations [20] that up to 10% of the power carried by one incident mode can be redistributed to other forward-propagating modes (see Fig. 1(c)). Backscattering with at most 0.1% is negligible. The general trend is that coupling is most efficient between adjacent modes (near the diagonal).

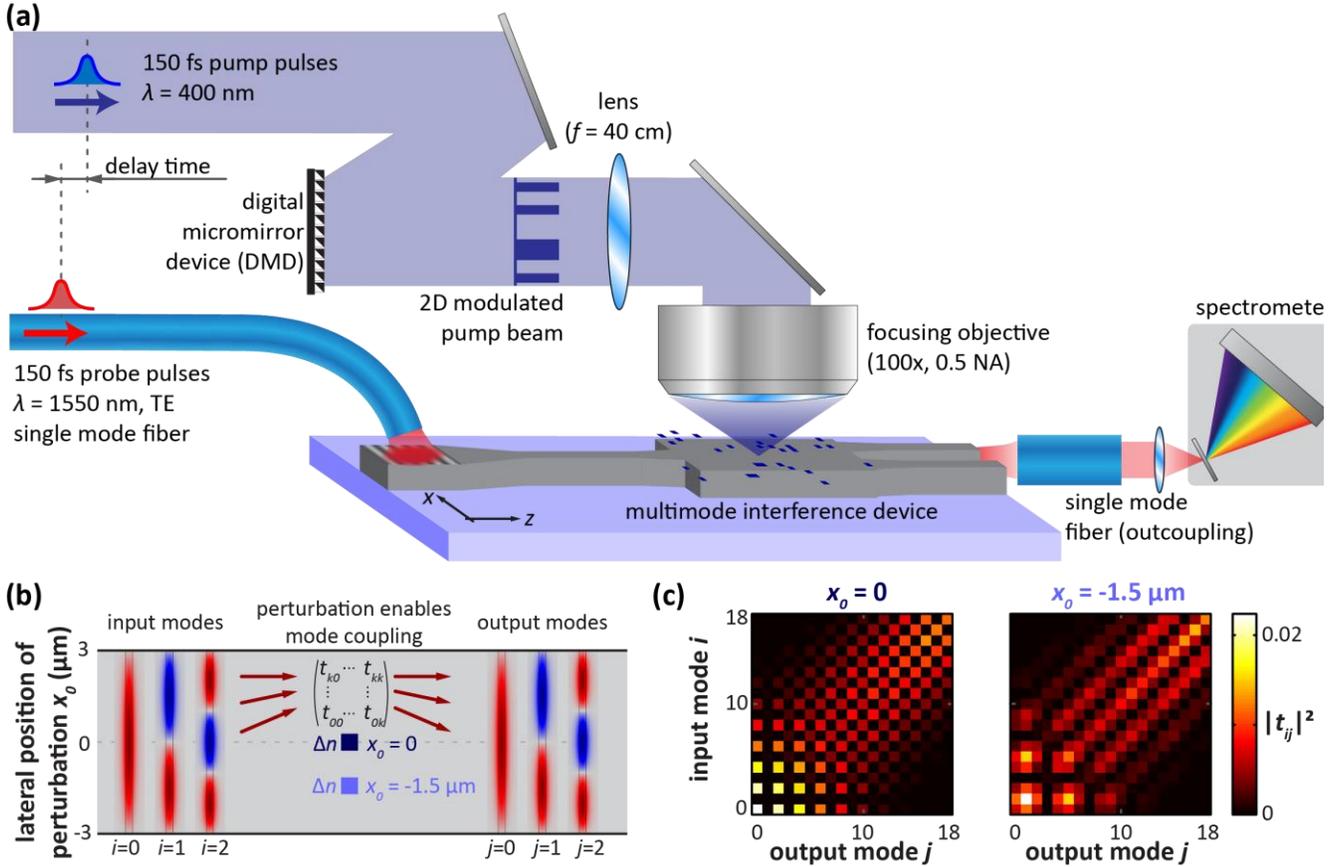

**Fig. 1, Concept of wavefront shaping by ultrafast photomodulation spectroscopy. (a)** In the experiments, transmission spectra of TE-polarized 150 fs probe pulses with a central wavelength of 1550 nm are monitored through a multimode interference (MMI) device. Simultaneously, a 2D pattern of 400 nm pump light is projected onto the device (blue overlay), locally decreasing the refractive index of the silicon MMI material by plasma dispersion. The pump beam is spatially modulated employing a digital micro-mirror device (DMD) and the pattern of the DMD is imaged onto the MMI surface by means of a lens and a microscope objective. **(b)** Each point of the pump pattern induces a perturbation in the refractive index profile of the MMI, enabling coupling between MMI modes, thus allowing shaping of the mode interference pattern at the MMI output plane. **(c)** Visual representation of the power mode coupling matrix for refractive index perturbations ($\Delta n = -0.25$, size 700 nm x 700 nm) at two lateral positions. Calculated from a-FMM simulations, these values correspond to the fraction of power that is exchanged between modes due to the perturbation.

As the coupling efficiency depends on the mode amplitudes at the lateral position $x_0$ of the perturbation, choosing $x_0$ allows defining how power is redistributed. For instance, perturbations at the MMI center couple only modes exhibiting the same symmetry, whilst off-center perturbations allow coupling between all modes. Additionally, every individual perturbation introduces a weak phase-shift (≈0.1 rd). Thus, a perturbation pattern, composed of tens of local perturbations, gives control over the distribution of power into the transverse modes as well as their phase relations, and may eventually change the intensity pattern at the output plane/port to any desired form. By addressing all degrees of freedom of the MMI-device, that is the mode power distribution and the mode phase relations, our technique grants total control over the transfer function of the device.

MMI-devices with tuneable power splitting ratios have been analyzed using the particular properties of self-images formed at intermediate distances inside the MMI [21-23]. Tuning of the refractive index using thermal or electro-optic elements positioned at these specific intermediate cross-sectional planes allows controlling the asymmetry of the self-images at the output using a single control parameter. While such multimode interference power splitters (MIPSs) are of technological interest, they provide only a limited range of tuning capabilities while requiring relatively large device footprints of several hundreds of micrometres in length [21-25]. Compared to the output-balancing characteristics of MIPS designs, wavefront shaping relies on a distributed phase-control over the entire volume of the MMI-device and offers *full* control over the transfer matrix in devices which are at least an order of magnitude reduced in length. The MMI-device (220 nm thick silicon slab of size 6 µm x 31.87 µm) under investigation is a standard component in state-of-the-art silicon photonics and was experimentally verified to have insertion losses below 1 dB [26].

**2. Theoretical analysis of the wavefront shaping mechanism**

To understand how a small set of local perturbations can redirect the propagating light towards any specific output port and to evaluate the approach for all-optical routing purposes, we resort to 2D coupled-mode simulations, performed with a fully-vectorial aperiodic-Fourier Modal Method (a-FMM) [20] (see Methods, $\lambda = $ 1550 nm, TE-polarization). Figure 2(a) (left) shows the electric-field map of the unperturbed MMI. Without fit parameters, simulations confirm the optimized MMI design with transmission exceeding 91% (< 0.5 dB loss) to the fundamental modes of the output waveguides (45.6% each). The on-axis input waveguide excites even (i.e. symmetric) MMI modes only, as shown in the left graph of Fig. 2(b). In an unperturbed MMI, modes would not exchange power and the mode power distribution would remain constant as light propagates through the MMI, resulting in a symmetric output profile. However, the inclusion of perturbations yields a redistribution of the incident power between MMI modes. The possibility to imprint desired modal phase and power distributions is the key to form a targeted intensity profile at the MMI output facet. For instance, maximizing light transmission to a single off-axis output port requires significant feeding of power to certain odd (i.e. anti-symmetric) modes. About 50% of the total power is required in odd modes, mostly in modes #1, #5, and #9. The right graph in Fig. 2(b) gives the target mode distribution needed for perfect overlap with the waveguide mode of the upper port. As the output waveguides feature 1 μm wide tapers, and thus a comparatively wide intensity profile, high-order MMI modes with high spatial-frequencies (#10 to #18) contribute only weakly to the interference pattern. Narrower waveguides would require stronger contributions from high-order modes.

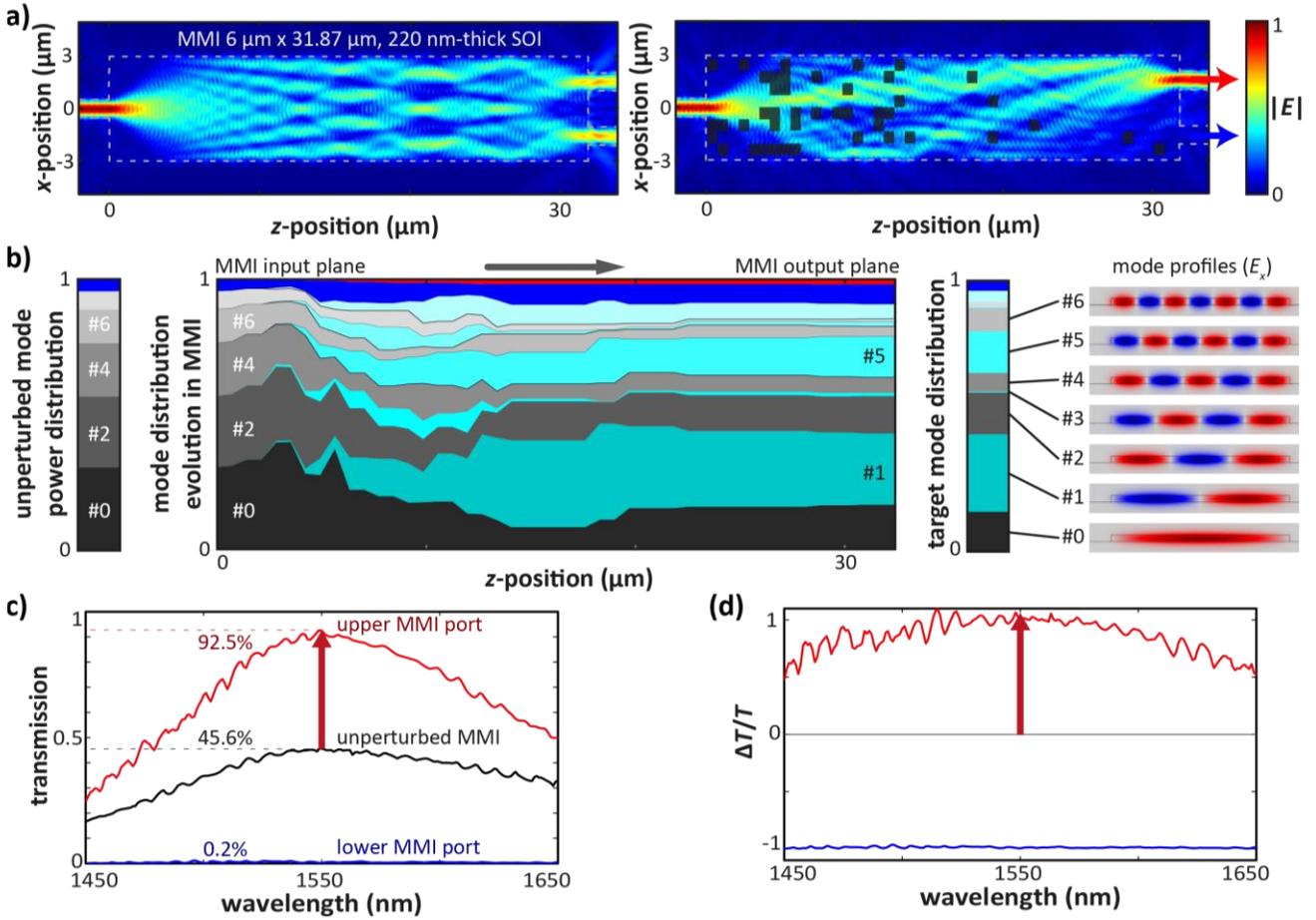

**Fig. 2, Numerical coupled-mode simulations. (a)** Electric field maps of the unperturbed MMI (left) and the MMI with a perturbation pattern with Δn = -0.25 (right), as indicated by the black overlay, from a-FMM simulations. The perturbation pattern was numerically optimized for maximum transmission into the upper MMI output port. **(b)** The injected light from the centrally placed input waveguide excites only even (symmetric) MMI modes at the input plane of the MMI (left graph). Due to the perturbation pattern, power is transferred from even modes (grey shades) into odd (anti-symmetric) modes (cyan shades) as the light propagates through the MMI (centre graph). The ten lowest order modes are plotted explicitly. For clarity, we combine the remaining nine MMI modes as they carry only little energy (blue area). Losses due to the perturbation pattern are shown in red. The resulting mode distribution interferes at the output plane of the MMI and steers the light effectively towards the upper output. We compare this mode distribution with the numerically ideal mode distribution (right graph) for maximum overlap with the output waveguide mode. The insets depict cross-sectional electric field profiles of the MMI modes calculated with Comsol Multiphysics. **(c,d)** Spectral behaviour of the beam steering effect produced by the optimized perturbation pattern. Transmissions into the fundamental modes of the top ($T_{top}$, red curve) and bottom ($T_{bot}$, blue curve) output waveguides remain > 85% and < 1% for a 60 nm bandwidth, respectively. The spectrum of the unperturbed MMI ($T_{ref}$) is given as black curve. The resulting transmission enhancement $\Delta T/T = (T_{top} - T_{ref})/T_{ref}$ and suppression $\Delta T/T = (T_{bot} - T_{ref})/T_{ref}$ are similarly broadband.

While single perturbations lack sufficient strength to excite asymmetric modes strongly, we show that a set of perturbations can be designed to collectively reach the desired requirement. To numerically optimize the perturbation pattern, we divide the MMI area into 700 nm-wide square pixels and use a random walk process to create a Markov chain in the pixel-state space. Considering that back-reflection is negligible, the random walk sequence should follow the direction of the light flow. Starting from the unperturbed MMI, the pixels are visited line-wise from the input plane towards the output plane. Within each line, pixels are addressed in a random sequence. For each visit, we calculate the transmission into the fundamental modes of the output waveguides with the a-FMM for an unperturbed ($\Delta n$ = 0) and a perturbed ($\Delta n$ = -0.25) pixel. The perturbation is kept in the set only if the transmission to the output desired port increases. Figure 2(a) (right) shows the electric field map of the MMI for an optimized perturbation pattern, while the centre graph in Fig. 2(b) concurrently visualizes the evolution of the MMI mode power distribution as light propagates through this perturbed MMI. The generated mode power distribution at the output plane of the MMI almost perfectly matches the ideal distribution (right graph). Losses (red) due to scattering into radiation modes and back-reflection are below 2%.

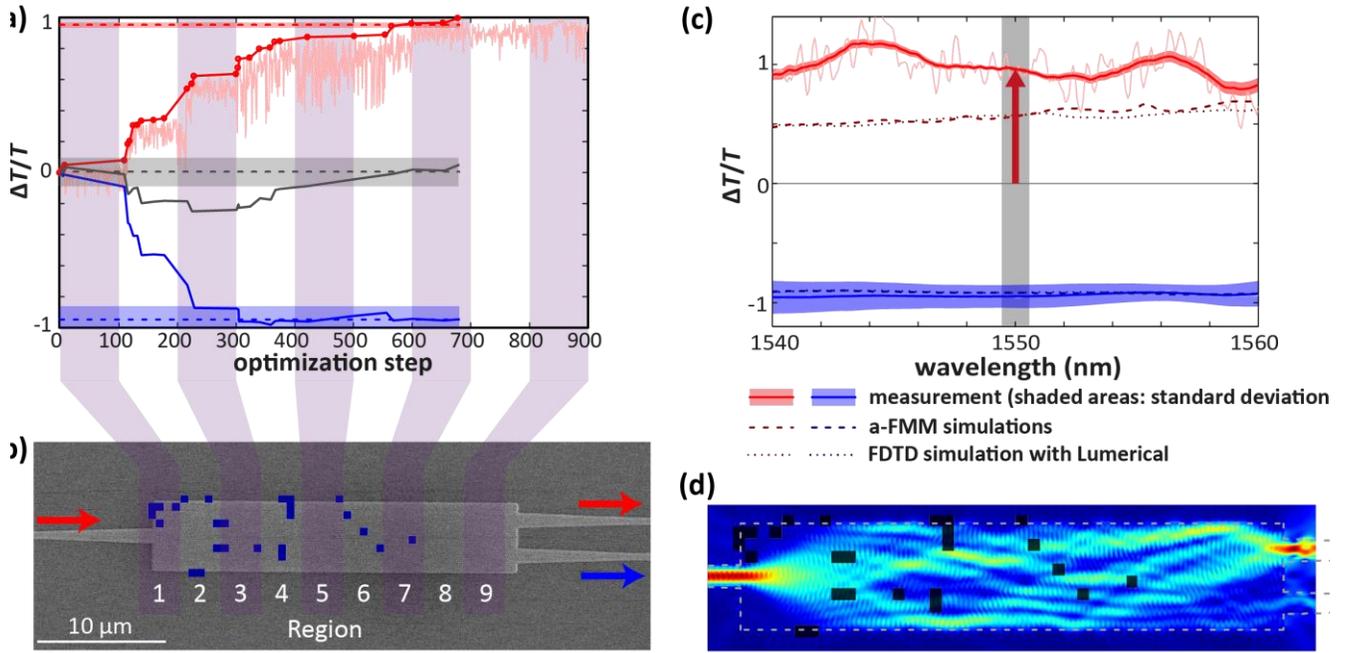

**Fig. 3, Digital micro-mirror device pattern optimization. (a)** Experimental optimization run to maximize transmission in a wavelength window of 1550 nm +/- 1.7 nm through the upper output port of the MMI. The individual regions (numbered from 1-9, 10 x 5 pixels each) are optimized individually in ascending order by performing 100 optimization steps in each region. In each step, a random pixel in the region under optimization is flipped, changing the transmission through the upper output port. The thin red line gives the change in transmission $\Delta T/T$ between the perturbed and unperturbed state for each step. Only pixels resulting in an increased $\Delta T/T$ are accepted (dots with thick red line) and the new DMD configuration is stored. In a subsequent measurement, all stored DMD configurations are reapplied while monitoring the lower MMI output (thick blue line). The excess losses induced in the MMI are calculated as the sum of $\Delta T/T$ of both outputs (thick grey line). After finishing the optimization run, we averaged 50 measurements with the final DMD configuration to determine the optimized $\Delta T/T$ for both ports minimally influenced by laser noise (horizontal dashed lines, shaded areas give the standard deviation). **(b)** SEM picture of the MMI-device overlaid with the final DMD configuration (blue) and indications of the regions used for the optimization. **(c)** $\Delta T/T$ spectra with the final DMD configuration for the two MMI outputs (thick solid lines; upper output red, lower output blue, shaded areas give the standard deviation), compared with simulated $\Delta T/T$ spectra from a-FMM (dashed lines) and FDTD method (dotted lines). The red arrow and the grey shaded area indicate the wavelength window used for optimization. **(d)** Simulated electric field maps from a-FMM ($\lambda$ = 1550 nm) for the MMI where for each pixel of the experimentally optimized pattern (black overlay) the effective index is reduced by 0.25.

The calculated transmission spectra of the MMI with the optimized perturbation pattern are shown in Fig. 2(c). The transmission of the unperturbed MMI (black) is increased in the upper port, with the transmission into the fundamental mode of the upper port reaching ~92.5% (-0.3 dB), surpassing the transmission of the unperturbed device (~91% in both ports) thanks to an enhanced mode matching at the output plane. Additionally, crosstalk is very weak with a transmission of only 0.2% (-27 dB) in the guided mode of the lower port. Expressed in terms of relative transmission changes ($\Delta T/T$), which may be measured experimentally more accurately than absolute transmission, $\Delta T/T$ reaches and exceeds values of 1 for the upper port and is getting close to -1 for the lower port (see Fig. 2(d)). It is interesting to note that the beam steering effect of a single perturbation pattern is effective on a broad wavelength range. The high transmission and very low cross-talk on a broad spectral range mark the excellent performance of the device. In Fig. (S1) of the Supplement Material, we further evidence that the excellent performance resist various experimental errors or uncertainties on the exact pattern position and perturbation strength.

### 3. Experimental realization of all-optical routing

The experimental demonstration of all-optical routing of light follows a similar scheme as the numerical calculations to optimize the digital pixel pattern of the DMD. For this purpose, we divided the MMI area into 9 regions of 10 vertical by 5 horizontal pixels and optimized each region in ascending order by performing 100 optimization steps (on average two visits per pixel are performed to reduce influence from laser noise).

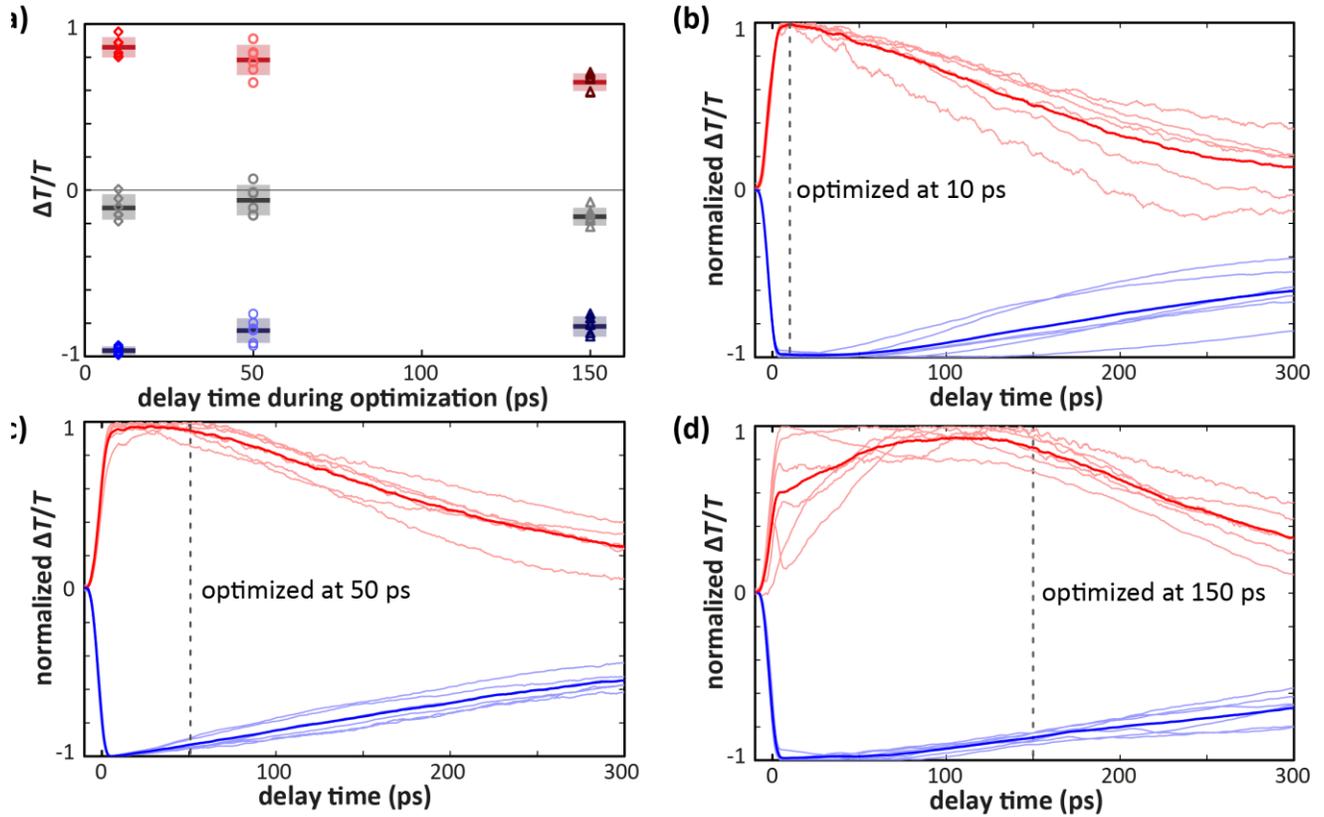

**Fig. 4, Time characteristics of excitation. (a)** $\Delta T/T$-values with DMD configurations after optimizations performed for three different delay times (10 ps - diamonds, 50 ps - circles, 150 ps - triangles, 6 values each). Thick horizontal lines give the group average and the shaded areas indicate the standard deviation. **(b-d)** Time characteristics induced by different pump light patterns (thin lines) optimized at different delay times (b: 10 ps, c: 50 ps, d: 150 ps). These curves are normalized to their individual $\Delta T/T$ maximum. Further, the averages (thick lines) of the 6-curve groups are given. We note that the averages are based on normalized curves, but are not normalized themselves. Thus, they do not necessarily reach a value of 1 or -1.

Figure 3(a) illustrates the experimental optimization process to maximize $\Delta T/T$ of the upper output of the MMI-device in a wavelength window of 1550 nm +/- 1.7 nm. The thin red curve gives the evolution of $\Delta T/T$ of the upper port at each optimization step. Every time a new transmission maximum is achieved, the DMD configuration is accepted (dots on the thick red line). After completing the optimization of the upper port, we analyze the transmission of the lower port in a subsequent experiment by reapplying all stored DMD configurations while measuring the lower-port transmission (thick blue line). We then calculate the induced excess losses (grey line) by our modulation technique as the sum of $\Delta T/T$ of both outputs. To minimize the influence of laser noise on our results, we averaged 50 measurements with the final DMD configuration for both outputs (horizontal dashed lines with shaded areas giving the standard deviation). The optimization shown in Fig. 3(a) was performed at a delay time between the pump and probe pulses of 10 ps, that is that pump pulses arrive at the MMI 10 ps before the probe pulses. For the final DMD configuration consisting of 25 active pixels, the transmission into the upper MMI output could be increased by more than 95% ($\Delta T/T = 0.96 \pm 0.02$), efficiently routing almost all (> 97.5%) transmitted light into one output. From $\Delta T/T = -0.95 \pm 0.09$ for the lower output, we deduct that excess losses are smaller than the accuracy of our experimental setup. Figure 3(b) gives a SEM micrograph of the MMI-device overlaid (blue) with the pump light pattern evolved in the optimization process. As in the simulations, we observe that the pixels are distributed over the MMI in a highly irregular pattern, showing the fundamental difference with MIPSs based on intermediate self-imaging conditions [21-25].

Figure 3(c) shows the final $\Delta T/T$ spectra after optimization (solid lines) and compares these with numerical results from the a-FMM (dashed lines) and finite-difference time-domain (FDTD) method (dotted lines) using the experimental perturbation pattern ($\Delta n = -0.25$). The resulting a-FMM electric field map in Fig. 3(d) confirms the beam steering effect, resulting in a field profile at the MMI output plane that matches the selected output waveguide. The difference in magnitude of the output enhancement between experimental and numerical spectra in Fig. 3(c) is due to the fact that simulations do not reflect experimental imperfections, such as inhomogeneity of the pump intensity over the DMD area, diffraction effects, and aberrations, which lead to deviations from the ideal pixel shape at the sample surface.

Nevertheless, it is important to note that the experimental optimization procedure naturally takes into account and "corrects" all these imperfections, resulting in enhancement levels comparable to those of the numerically optimized MMI shown in Fig. 2. A strong enhancement of the transmission is achieved experimentally over the entire observed 20 nm wavelength range, demonstrating the broadband characteristics of the integrated spatial light modulator. Similarly to the numerical results, $\Delta T/T$ values larger than 1 are possible as the optimization can result in an increase of the overall device transmission compared to the unperturbed state. Averaged over the whole 20 nm wide wavelength window, we find an average ΔT/T in the upper waveguide of 0.98 ± 0.1, and -0.94 ± 0.09 in the lower waveguide. The sensitivity of the pump pattern alignment was investigated both theoretically and experimentally in Fig. (S1) and (S2) of the Supplementary Material. It is found that perturbation patterns have alignment tolerances of less than 1 µm in both the propagation and transverse directions. This sensitivity reveals the delicate coupling between many degrees of freedom in a complex system, with an output pattern that is the result of many subsequent perturbations acting in concert.

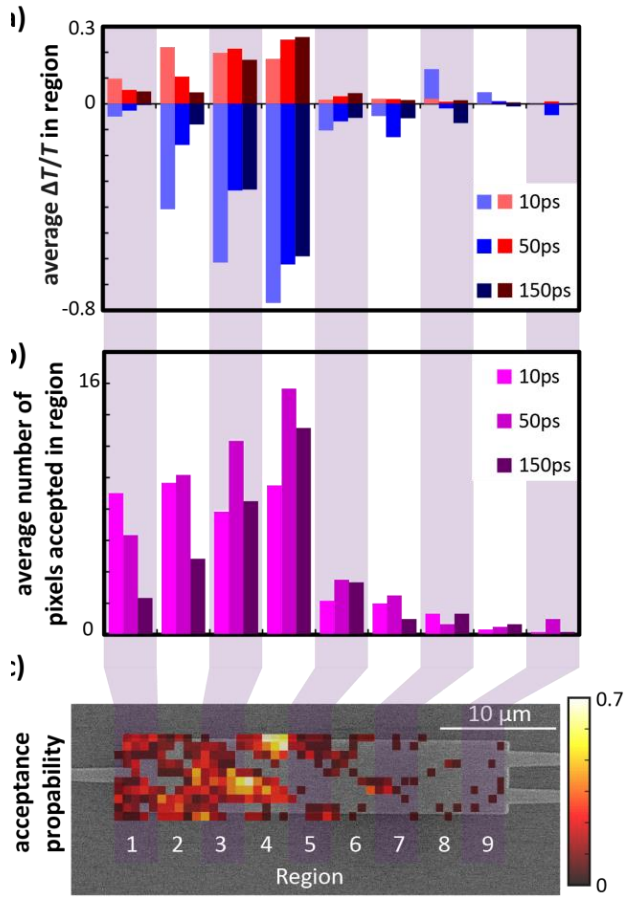

**Fig. 5, Contribution of individual MMI regions. (a)** Average ΔT/T per region in the experiments presented in Fig. 4. The total ΔT/T achieved during the optimization is the sum of ΔT/T in the individual regions. **(b)** Average numbers of pixels accepted per region during the optimization processes. **(c)** Map of the probability that individual pixels are accepted during optimization.

The spatial optimization is the result of a series of local perturbation that depend on time as the system evolves from excited states induced by the ultrafast pump pulses. To investigate the role of time dynamics, perturbation pattern were optimized for different delay times (10 ps, 50 ps, 150 ps), with the results summarized in Fig. 4(a). For longer delay times, the probe light experiences smaller refractive index perturbations as the free carriers exited by the pump are given more time to recombine (see Fig. (S3) in the Supplementary Material). As a result, the largest increase in transmission could be obtained for the shortest delay time of 10 ps, where the average ΔT/T = 0.86 ± 0.06, with the average residual light transmission in the other output under 4%. At longer delay times, maximum enhancements are reduced. However, even at 150 ps, which is of the order of the free carrier lifetime of 174 ps, transmission increases exceeding 65% can be achieved. The time characteristics of the optimized devices are analyzed in Figs. 4(b-d). As expected, the maximum effect is mostly achieved around the delay time at which the optimization was performed. Furthermore, configurations optimized at 10 ps show the fastest decay with a recovery time of around 190 ps. Optimizations at longer delay times result in slower initial rise and recovery times of the device, with a 100 ps-long plateau of nearly constant performance.

Different optimization runs generally resulted in individual patterns because of the randomness introduced in the iterative process, and due to intrinsic noise present in the experimental system. However, general trends could be observed in the relative contributions of different parts of the MMI. In Fig. 5(a), the average ΔT/T obtained per region is revealed for the set of 18 experiments of Fig. 4, while Fig. 5(b) gives the average number of pixels accepted. Clearly, regions 2 to 4 are the most active and contribute strongly to the optimization process, with the most increase in transmission and the most pixels accepted. On average, 43.8 pixels are accepted per run, representing less than 10% of the MMI area and showing that a relatively small number of perturbations is sufficient to efficiently shape wavefronts in silicon devices. In the two dimensional map displayed in Fig. 5(c), the acceptance probability for each pixel position is given for an optimization of the upper output.

Wavefront shaping in silicon photonics holds potential for a wide range of device applications. We foresee many ways for further optimization of this concept, for example using the additional degrees of freedom in high-resolution DMD spatial light modulators to develop gradient designs based on transformation optics, or the use of phase change layers for writing non-volatile patterns for reconfigurable optical memory devices. Once optimized patterns have been obtained using the relatively slow iterative process, they can be stored in a library, or imprinted using other techniques such as amplitude masks or multiple beam steering and applied at high frequencies. In addition to optimizing spatial degrees of freedom in broadband devices, it will be possible to design optical systems with wavelength dependent responses and to use wavefront shaping to achieve spectral control, or specific time domain characteristics, similar to their three-dimensional counterparts.

## 4. Conclusions

In summary, we showed that arbitrary high-contrast (Δn = -0.25) binary refractive index profiles can be generated in planar silicon exploiting plasma dispersion from all-optical excitation with spatially modulated pump beams. We experimentally realized dynamic routing of light in a 1x2 MMI power splitter to any output port with better than 97% efficiency, cross talk below -27 dB, and negligible losses. Ultimately limited by the free carrier recombination time in the MMI, switching speeds in the gigahertz regime could potentially be reached.

Our theoretical considerations indicate that by addressing the mode power distribution and the mode phase relations, our technique can exercise full control over the device transmission function and generate arbitrary intensity pattern at the output facet of the device. Thus, we turn one of the most common passive optical elements into a versatile platform for reconfigurable silicon photonics.

We believe that the present integrated distributed control is a new and effective approach for parallel dynamic control of degrees of freedom and that the presented concept can be easily extended for more complex devices, for instance with several output ports. In addition, the present distributed approach can be viewed as the integrated-optics analogue of volume holograms and thus may benefit from the additional degrees of freedom of volume holography [27]. This new concept of integrated spatial light modulation in silicon thus opens up avenues to all-optical reconfigurable devices with possible applications in testing of optical circuits and reconfigurable multi-port optical filters, splitters, and modulators for data-communication.

**Dataset citation.** The dataset for this paper can be found at http://dx.doi.org/10.5258/SOTON/384788.

**Funding**. The authors acknowledge support from EPSRC through grant no. EP/J016918. Goran Z. Mashanovich acknowledges support from the Royal Society through a University Research Fellowship.

See Supplement 1 for supporting content.

**Methods**

**Optical setup.** The laser system used is an amplified 800 nm Ti-Sapphire laser coupled with an optical parametric amplifier (OPA) providing 150 fs pulses at a 250 kHz repetition rate. The system gives a frequency doubled output at 400 nm, used as the pump, and an idler output of the parametric amplifier which was tuned to 1550 nm for the presented measurements. In our setup (see Fig. 1(a)) the probe pulses are in-coupled by the use of a grating coupler from a singlemode fibre, using TE polarization. The transmitted light of one output of the MMI-device is end-face coupled into a singlemode fibre and subsequently to a spectrometer. The pump beam is expanded to a diameter of about 15 mm and spatially modulated by a Texas Instruments DLP7000 digital micro-mirror device (DMD) with a resolution of 1024 x 768 physical pixels and a pixel pitch of 13.6 µm. For our experiments, we combine 10 x 10 physical pixels of the DMD to one logical pixel (see below). The main text addresses logical pixels only. The imaging system comprises a 40 cm lens and a 100× 0.5NA Mitutoyo microscope objective, to reduce the pattern size from the DMD by a factor 200, thus translating the 136 µm edge length logical pixels into spots of about 700 nm on the MMI-device surface. We note that the 18 experiments presented in Figs. 3-5 have not been preselected and represent a set of 18 consecutive experiments. For these experiments the output port for which $\Delta T/T$ was maximized was alternated to prove the symmetry of the device and our approach. In the analysis, the change in transmission $\Delta T/T$ is used rather than the absolute transmission to correct for inherent experimental variations in laser intensity and for differences in the throughput of the individual output waveguides and cleaved end facets. A complete experimental optimization process requires around 15 minutes.

**MMI-device.** The MMI 1×2 power splitter was optimized using 3D FIMMWAVE simulations and fabricated from a 220 nm-thick silicon layer on top of a 2 µm-thick $SiO_2$ layer [26]. The MMI region has a design length of 31.875 µm and a width of 6 µm. The separation of the two MMI output waveguides is 3.14 µm (center-to-center). Single mode 400 nm-wide waveguides are tapered to 1 µm at the MMI boundaries, ensuring an adiabatic size conversion of the fundamental mode over a 10 µm-distance.

**Refractive index modulation.** In our experiments, we measured an average pump fluence of 64 pJ/µm² incident onto the MMI surface. Based on our previous study [16], this changes the real part of the refractive index by -0.25 in the pumped area. Larger refractive index modulations up to -0.5 are possible without damaging the silicon. However, in the presented experiments, we were limited by the total pump power provided by the laser system, which has to be delivered over the whole area of the MMI.

**Fully-vectorial modal simulations.** The simulations presented in Figs. 1-3, and in the Supplementary Material, were performed using an in-house fully-vectorial frequency-domain Fourier Modal method (a-FMM) [20]. This method calculates the modes in a Fourier basis, using a supercell method and perfectly-matched layers. It relies on a scattering-matrix formalism to describe accurately the coupling between modes propagating in the input waveguide, MMI, and output waveguides. Simulations were performed in 2D with a wavelength-dependent effective index (2.85 at a wavelength of 1.55 µm) to account for the vertical confinement of the fundamental TE-mode of the 220 nm-thick silicon slab. Out-of-plane scattering produced by the refractive index modulations in the cladding is expected to be weak due to the small refractive index change ($\Delta n$ = -0.25), thus we expect that 2D simulations can accurately predict the device performance. One simulation typically takes 1-3 seconds with a personal computer, thereby allowing rapid investigations of MMIs. MMI design parameters were used for the simulations. To reproduce the experimental beam steering effect presented in Fig. 3, we performed a parametric study of the device performance, optimizing the lateral and longitudinal position of the experimental perturbation pattern. The pattern position was therefore the only free parameter.

**Experimental optimization process.** *Influence of laser shot noise.* During experimental optimisation processes, $\Delta T/T$ before and after flipping a pixel can only be determined with certain accuracy due to laser shot noise. This unavoidably leads to the acceptance of a small number of pixels at the beginning of the optimization process that do not physically contribute to the wavefront shaping process. We counteract this fact by positioning region 1 partly outside the MMI region allowing pixels being accepted due to laser noise without affecting the wavefront.

*Choice of logical pixel size.* In the presented experiments we combined 10 x 10 physical pixels of the DMD to one logical pixel. In theory, smaller logical pixels would allow for a finer control over the mode distribution. However, by reducing pixel size, the impact of individual pixels on $\Delta T/T$ is smaller and thus more difficult to detect in the presence of laser noise. We experimented with different logical pixel sizes (5 x 5 to 20 x 20 physical pixels), with 10 x 10 physical pixels giving the most reliable results without sacrificing device performance.

*Advanced algorithms.* To further improve the quality of wavefront shaping, we evaluated more advanced optimization algorithms in the experiments. In a first approach, we applied the presented algorithm twice, using the end configuration of the first run as start configuration of the second run. This approach yielded only negligible return. Additionally, we investigated genetic algorithms, where DMD configurations found with the presented algorithm (parent-configurations) were combined and mutated to create new configurations (child-configurations). However, no significant improvement could be achieved due to the nature of the optimization problem and limitations in the experimental setup. Firstly, the optimization function is of high dimension and generally has many local maxima, that is that many different DMD configurations can satisfy a large $\Delta T/T$ increase. However, the same is not necessarily true for combinations from these patterns, strongly reducing the probability that child-configurations surpass their parent-configurations. Secondly, starting from already excellent results with the algorithm presented, only small improvements are expected, which are difficult to detect due to laser shot noise and require long averaging periods, eventually increasing the optimization time beyond a point where the long-term stability of the used laser system can be guaranteed.

**Running average on presented spectra.** The spectra presented in Fig. 3(c) show a ripple structure with a period of about 0.6 nm, which is consistent with reflections from the cleaved end face at the end of the output waveguides. Hence we plot the measured spectra (thin lines) overlaid with a running average (thick lines), with an averaging window of +/- 2.1 nm (+/- 25 spectral points provided by the spectrometer).